\documentclass[12pt]{article}
\usepackage{graphicx}
\usepackage[cp1251]{inputenc}
\usepackage[russian]{babel}
\usepackage{rotating}
\begin{document}
 \noindent {\footnotesize\it Astronomy Reports, 2017, Vol. 61, No 7, pp. 551--559}
 \newcommand{\dif}{\textrm{d}}

 \noindent
 \begin{tabular}{llllllllllllllllllllllllllllllllllllllllllllll}
 & & & & & & & & & & & & & & & & & & & & & & & & & & & & & & & & & & & & & \\\hline\hline
 \end{tabular}

  \vskip 0.5cm
 \centerline{\bf\large Galactic Orbits of Globular Clusters in the Region}
 \centerline{\bf\large of the Galactic Bulge}
 \bigskip
 \bigskip
  \centerline
 {
 V.V. Bobylev,
 A.T. Bajkova
 }
 \bigskip
{\small \it
Central (Pulkovo) Astronomical Observatory, Russian
Academy of Sciences, Pulkovskoe shosse 65, St. Petersburg, 196140
Russia
 }
 \bigskip
 \bigskip
 \bigskip

 {
{\bf Abstract}---Galactic orbits have been constructed over long
time intervals for ten globular clusters located near the Galactic
center. A model with an axially symmetric gravitational potential
for the Galaxy was initially applied, after which a non-axially
symmetric potential corresponding to the central bar was added.
Variations in the trajectories of all these globular clusters in
the $XY$ plane due to the influence of the bar were detected.
These were greatest for the cluster Terzan~4 in the meridional
$(RZ)$ plane. The globular clusters Terzan~1, Terzan~2, Terzan~4,
Terzan 9, NGC 6522, and NGC 6558 always remained within the
Galactic bulge, no farther than 4 kpc from the Galactic center.
  }

 \section{INTRODUCTION}
Globular clusters are an important source of information for
studies of the structure and evolution of the Galaxy [1,2].
Distant globular clusters are of considerable interest for studies
of the properties of the halo, the distribution of the matter
density in the Galaxy, and estimates of its mass [3--5]. For
example, globular clusters located near the Galactic center [6]
have been used to investigate the properties of the bulge and bar
[7,8] or star-formation processes after the passage of a globular
cluster through the disk [9].

Many globular clusters in the Milky Way have measured absolute
proper motions. For example, the proper motions of globular
clusters were determined in [10--13] based on ground photographic
and CCD observations of stars in the Southern hemisphere [14],
combined with data from modern HIPPARCOS catalogs [15] to provide
second-epoch observations. These data were used in [7,8] to
construct Galactic orbits for more than 50 globular clusters.
Proper motions for 92 globular clusters were determined in [16]
using data from the UCAC2 catalog [17]. Absolute proper motions of
more than 140 globular clusters were computed in [18] using data
from the PPMXL catalog [19]. The UCAC2 and PPMXL catalogs contain
absolute proper motions of stars. They extend the HIPPARCOS system
to weaker stars, but contain significant errors, in the brightness
equation for UCAC2 and in the form of zonal inhomogeneities of
order 2 milliarcsecond/year (2 mas/yr) for PPMXL, as was shown for
example in [20] and [21].

When studying the kinematics of Galactic globular clusters, most
of which are far from the Sun, it is of most interest to use
proper motions obtained using the largest ground telescopes at
appreciably different epochs, or with the Hubble Space Telescope
(HST), where uncertainties in the measurements and absolute values
are determined directly using images of distant galaxies. We have
made use of precisely such observations in our current study.

The construction of Galactic orbits for globular clusters requires
a good model for the gravitational potential of the Galaxy. In
[22], we refined the parameters of three models for the Galactic
potential with different forms for the dark-matter halo. We used
modern sets of observational data encompassing a wide range of
distances to the rotation axis $R,$ from0 to $\sim$200 kpc.
Moreover, it was shown in [7,8] that the central bar of the Galaxy
appreciably influences the motions of globular clusters.
Therefore, it is especially important to take this influence into
account when analyzing the motions of globular clusters located
near the bulge and bar. Our goal in the current study was to
investigate the three-dimensional kinematics of globular clusters
in the Milky Way using measured distances, radial velocities, and
proper motions. We constructed their Galactic orbits using a
refined model for the gravitational potential of the Galaxy. The
orbits were constructed for two cases: using an axially symmetric
potential, and adding the non-axially symmetric potential of the
bar.

 \section{METHOD}
 \subsection{Model for the Galactic Potential}
The axially symmetric gravitational potential of the Galaxy was
represented as the sum of three components --- the central,
spherical bulge $\Phi_b(r(R,Z))$, the disk $\Phi_d(r(R,Z))$, and
the massive, spherical dark-matter halo $\Phi_h(r(R,Z))$:
 \begin{equation}
 \begin{array}{lll}
  \Phi(R,Z)=\Phi_b(r(R,Z))+\Phi_d(r(R,Z))+\Phi_h(r(R,Z)).
 \label{pot}
 \end{array}
 \end{equation}
Here, we used a cylindrical coordinate system ($R,\psi,Z$) with
its origin at the Galactic center. In Cartesian coordinates
$(X,Y,Z)$ with their origin at the Galactic center, the distance
to a star (the spherical radius) is $r^2=X^2+Y^2+Z^2=R^2+Z^2,$
where the $X$ axis is directed from the Sun toward the Galactic
center, the Y axis is perpendicular to the $X$ axis and points in
the direction of the Galactic rotation, and the $Z$ axis is
perpendicular to the Galactic $(XY)$ plane and points in the
direction of the North Galactic pole. The gravitational potential
is expressed in units of 100 km$^2$ s$^{-2}$, distances in kpc,
masses in units of the mass of the Galaxy, $M_{gal}=2.325\times
10^7 M_\odot$, and the gravitational constant is taken to be
$G=1.$

The potentials of the bulge $\Phi_b(r(R,Z))$ and disk
$\Phi_d(r(R,Z))$ were taken to have the form proposed by Miyamoto
and Nagai [23]:
 \begin{equation}
  \Phi_b(r)=-\frac{M_b}{(r^2+b_b^2)^{1/2}},
  \label{bulge}
 \end{equation}
 \begin{equation}
 \Phi_d(R,Z)=-\frac{M_d}{\Biggl[R^2+\Bigl(a_d+\sqrt{Z^2+b_d^2}\Bigr)^2\Biggr]^{1/2}},
 \label{disk}
\end{equation}
where $M_b$ and $M_d$ are the masses of the corresponding
components and $b_b, a_d,$ and $b_d$ are scale parameters of the
components in kpc. According to [24], the halo component can be
represented
 \begin{equation}
  \Phi_h(r)=-\frac{M_h}{r} \ln {\Biggl(1+\frac{r}{a_h}\Biggr)}.
 \label{halo-III}
 \end{equation}
Table 1 presents the parameters of the model for the Galactic
potential (2)--(4) derived by Bajkova and Bobylev [22] using a
Galactic rotation curve constructed for objects located at
distances $R$ out to $\sim$200 kpc. The local parameter values
$R_\odot=8.3$ kpc and $V_\odot=244$ km s$^{-1}$ were used when
constructing this rotation curve. The model (2)--(4) was denoted
Model III in [22].

We chose to describe the potential of the central bar using the
triaxial ellipsoid model [25]:
\begin{equation}
  \Phi_{bar}=-\frac{M_{bar}}{(q_b^2+X^2+[Ya_b/b_b]^2+[Za_b/c_b]^2)^{1/2}},
\label{bar}
\end{equation}
where $X=R\cos\vartheta, Y=R\sin\vartheta$, $a_b, b_b, c_b$ are
the three semi-axes of the bar; $q_b$ is the length of the bar;
$\vartheta=\theta-\Omega_{bar}t-\theta_{bar}$, $tg(\theta)=Y/X$,
$\Omega_{bar}$ is the angular speed of the bar; $t$ is the
integration time; and $\theta_{bar}$ is the inclination of the bar
relative to the $X$ and $Y$ axes, measured from the line joining
the Sun and the Galactic center (the $X$ axis) to the major axis
of the bar in the direction of the Galactic rotation. We adopted
the angular speed of the bar $\Omega_{bar}=55$ km s$^{-1}$
kpc$^{-1}$, in accordance with the estimates of Bobylev and
Bajkova [26].

%%%%%%%%%%%%%%%%%%%%%%%%%%%%%%%%%%%%%%%%%%%%%%%%%%%%%%%%%%%%%%%%%%%
 {\begin{table}[t]                                    %% Т~1.
 \caption[]
 {\small\baselineskip=1.0ex
Parameters of Model III for the Galactic potential, according to
[22], with $M_{gal}=2.325\times 10^7 M_\odot$
  }
 \label{t:model-III}
 \begin{center}\begin{tabular}{|c|r|}\hline
 $M_b$ &   443 M$_{gal}$ \\
 $M_d$ &  2798 M$_{gal}$ \\
 $M_h$ & 12474 M$_{gal}$ \\
 $b_b$ & 0.2672 kpc  \\
 $a_d$ &   4.40 kpc  \\
 $b_d$ & 0.3084 kpc  \\
 $a_h$ &    7.7 kpc  \\
 $M_{bar}$ & 43.1 M$_{gal}$ \\
 $q_b$     &  5.0 kpc  \\
 $a_b/b_b$ & $1/0.42$  \\
 $a_b/c_b$ & $1/0.33$  \\\hline
 \end{tabular}\end{center}\end{table}}
%%%%%%%%%%%%%%%%%%%%%%%%%%%%%%%%%%%%%%%%%%%%%%

 \subsection{Construction of the Orbits}
The equation of motion of a test particle in an axially symmetric
gravitational potential can be obtained from the Lagrangian
$\pounds$ of the system (see Appendix A of [27]):
\begin{equation}
 \begin{array}{lll}
 \pounds(R,Z,\dot{R},\dot{\psi},\dot{Z})=0.5(\dot{R}^2+(R\dot{\psi})^2+\dot{Z}^2)-\Phi(R,Z).
 \label{Lagr}
 \end{array}
\end{equation}
Introducing the canonical momenta
$p_{R}=\partial\pounds/\partial\dot{R}=\dot{R}$,
 $p_{\psi}=\partial\pounds/\partial\dot{\phi}=R^2\dot{\psi}$, and
 $p_{Z}=\partial\pounds/\partial\dot{Z}=\dot{Z}$,
we obtain the Lagrangian equations in the form of a system of six
first-order differential equations:
 \begin{equation}
 \begin{array}{llllll}
 \dot{R}=p_R,\\
 \dot{\psi}=p_{\psi}/R^2,\\
 \dot{Z}=p_Z,\\
 \dot{p_R}=-\partial\Phi(R,Z)/\partial R +p_{\psi}^2/R^3,\\
 \dot{p_{\psi}}=0,\\
 \dot{p_Z}=-\partial\Phi(R,Z)/\partial Z.
 \label{eq-motion}
 \end{array}
\end{equation}
We integrated Eqs. (7) using a fourth-order Runge--Kutta
algorithm.

The peculiar velocity of the Sun relative to the Local Standard of
Rest (LSR) was taken to be
$(u_\odot,v_\odot,w_\odot)=(11.1,12.2,7.3)\pm(0.7,0.5,0.4)$ km
s$^{-1}$, in accordance with [28]. Here, the heliocentric velocity
is given in a moving Cartesian coordinate system with $u$ directed
toward the Galactic center, $v$ in the direction of the Galactic
rotation, and $w$ perpendicular to the Galactic plane toward the
North Galactic pole.

We denoted the initial values of the positions and space
velocities of a test particle in the heliocentric coordinate
system $(x_o,y_o,z_o,u_o,v_o,w_o)$. These initial positions and
velocities are then given in the fixed Cartesian coordinates of
the Galactic system by the formulas
\begin{equation}
 \begin{array}{llllll}
 X=R_0-x_o, Y=y_o, Z=z_o,\\
 U=u_o+u_\odot,\\
 V=v_o+v_\odot+V_0,\\
 W=w_o+w_\odot,\\
 \Pi=-U\cos\psi_o+V\sin\psi_o,\\
 \Theta=U\sin\psi_o+V\cos\psi_o,
 \label{init}
 \end{array}
\end{equation}
where $R_0$ and $V_0$ are the Galactocentric distance and linear
velocity of the LSR about the center of the Galaxy and tan
$\tan\psi_o=Y/X$.

%%%%%%%%%%%%%%%%%%%%%%%%%%%%%%%%%%%%%%%%%%%%%%%%%%%%%%%%
 {\begin{table}[t]                                    %% T~2.
 \caption[]
 {\small\baselineskip=1.0ex
 Input data on the globular clusters according to [6]
  }
 \label{t:1}
 \small\baselineskip=1.0ex
\begin{center}\begin{tabular}{|l|r|r|r|r|r|r|r|}\hline
 Cluster   & $l,$ & $b,$ & $\mu_\alpha\cos\delta,$ & $\mu_\delta,$ & $V_r,$ & $d,$ \\
           & deg & deg & mas yr$^{-1}$ & mas yr$^{-1}$ & km s$^{-1}$ & kpc \\\hline
 Terzan 1  & $357.57$ & $  1.00$ & $ 0.51\pm0.31$ & $-0.93\pm0.29$ & $  114\pm14~ $ & $6.2\pm0.6$ \\ %  mu-Rossi2015
 Terzan 2  & $356.32$ & $  2.30$ & $-0.94\pm0.30$ & $ 0.15\pm0.42$ & $  109\pm15~ $ & $8.7\pm0.8$ \\ %  mu-Rossi2015
 Terzan 4  & $356.02$ & $  1.31$ & $ 3.50\pm0.69$ & $ 0.35\pm0.58$ & $ -50.0\pm2.9$ & $9.1\pm0.9$ \\ %  mu-Rossi2015
 Terzan 9  & $  3.61$ & $ -1.99$ & $ 0.00\pm0.38$ & $-3.07\pm0.49$ & $   59\pm10~ $ & $7.7\pm0.7$ \\ %  mu-Rossi2015
 NGC 6522  & $  1.02$ & $ -3.93$ & $ 3.35\pm0.60$ & $-1.19\pm0.34$ & $ -21.1\pm3.4$ & $7.8\pm0.7$ \\ %  mu-Rossi2015
 NGC 6540  & $  3.29$ & $ -3.31$ & $ 0.07\pm0.40$ & $ 1.90\pm0.57$ & $ -17.7\pm1.4$ & $3.7\pm0.3$ \\ %  mu-Rossi2015
 NGC 6558  & $  0.20$ & $ -6.02$ & $-0.12\pm0.55$ & $ 0.47\pm0.60$ & $-197.2\pm1.5$ & $7.4\pm0.7$ \\ %  mu-Rossi2015
 NGC 6652  & $  1.53$ & $-11.38$ & $ 4.75\pm0.07$ & $-4.45\pm0.10$ & $-111.7\pm5.8$ & $9.6\pm0.9$ \\ %  mu-Rossi2015
 NGC 6681  & $  1.53$ & $-12.51$ & $ 1.58\pm0.18$ & $-4.57\pm0.16$ & $ 220.3\pm0.9$ & $9.0\pm1.8$ \\ %  mu-Rossi2015
 Palomar 6 & $  2.10$ & $  1.78$ & $ 2.95\pm0.41$ & $ 1.24\pm0.19$ & $ 181.0\pm2.8$ & $7.3\pm0.7$ \\ %  mu-Rossi2015
 \hline
 \end{tabular}\end{center}\end{table}}
%%%%%%%%%%%%%%%%%%%%%%%%%%%%%%%%%%%%%%%%%%%%%%%%%%%%%%%%
%%%%%%%%%%%%%%%%%%%%%%%%%%%%%%%%%%%%%%%%%%%%%%%%%%%%%%%%
 {\begin{table}[t]                                    %% T~3.
 \caption[]
 {\small\baselineskip=1.0ex
Initial velocities in the fixed Cartesian coordinates $U,V,W$ and
the cylindrical coordinates $\Pi,\Theta$
  }
 \label{t:UVW}
 \small\baselineskip=1.0ex
\begin{center}\begin{tabular}{|l|r|r|r|r|r|r|r|}\hline
 Cluster & $U,$ kpc & $V,$ kpc & $W,$ kpc & $\Pi,$ kpc & $\Theta,$ kpc \\\hline
 Terzan 1  & $ 125\pm14$ & $236\pm12$ & $ -18\pm4~$ & $-153\pm14$ & $~219\pm12$ \\
 Terzan 2  & $ 117\pm14$ & $233\pm20$ & $  47\pm4~$ & $-128\pm19$ & $-227\pm16$ \\
 Terzan 4  & $ -30\pm1~$ & $355\pm38$ & $-112\pm16$ & $-247\pm24$ & $-257\pm30$ \\
 Terzan 9  & $  74\pm11$ & $163\pm25$ & $ -50\pm6~$ & $~~41\pm18$ & $~174\pm21$ \\
 NGC 6522  & $ -19\pm3~$ & $278\pm22$ & $-121\pm17$ & $~~91\pm6~$ & $~263\pm21$ \\
 NGC 6540  & $  -7\pm2~$ & $285\pm12$ & $  23\pm2~$ & $~~20\pm2~$ & $~284\pm12$ \\
 NGC 6558  & $-184\pm1~$ & $268\pm28$ & $  39\pm7~$ & $~191\pm1~$ & $~263\pm28$ \\
 NGC 6652  & $-151\pm1~$ & $160\pm10$ & $-247\pm25$ & $-112\pm2~$ & $-190\pm10$ \\
 NGC 6681  & $ 203\pm4~$ & $116\pm31$ & $-179\pm28$ & $~227\pm21$ & $~~52\pm23$ \\
 Palomar 6 & $ 191\pm4~$ & $353\pm16$ & $ -52\pm11$ & $~-94\pm5~$ & $~390\pm15$ \\
 \hline
 \end{tabular}\end{center}\end{table}}
%%%%%%%%%%%%%%%%%%%%%%%%%%%%%%%%%%%%%%%%%%%%%%%%%%%%%%%%
%%%%%%%%%%%%%%%%%%%%%%%%%%%%%%%%%%%%%%%%%%%%%%%%%%%%%%%%
 {\begin{table}[t]                                    %% T~4.
 \caption[]
 {\small\baselineskip=1.0ex
Characteristics of the globular-cluster orbits computed for the
axially symmetric potential (upper rows) and with the addition of
the bar potential (lower rows)
 }
 \label{t:ecc}
 \small\baselineskip=1.0ex
\begin{center}\begin{tabular}{|l|c|c|c|c|c|c|c|}\hline
 Cluster & $ a_{min}$ & $ a_{max}$ & $ e$ \\\hline
           & \multicolumn{2}{c|}{kpc} &     \\\hline
 Terzan 1  & 1.41 & 3.80 & 0.46 \\
 Terzan 2  & 0.51 & 1.15 & 0.39 \\
 Terzan 4  & 0.60 & 4.15 & 0.75 \\
 Terzan 9  & 0.42 & 0.89 & 0.36 \\
 NGC 6522  & 0.50 & 1.55 & 0.51 \\
 NGC 6540  & 4.58 & 6.70 & 0.19 \\
 NGC 6558  & 0.70 & 3.82 & 0.69 \\
 NGC 6652  & 1.64 & 5.83 & 0.56 \\
 NGC 6681  & 0.67 & 4.69 & 0.75 \\
 Palomar 6 & 1.00 & 5.29 & 0.68 \\ \hline

 Terzan 1  & 1.42 & 3.85 & 0.46 \\
 Terzan 2  & 0.52 & 1.21 & 0.40 \\
 Terzan 4  & 0.59 & 4.27 & 0.76 \\
 Terzan 9  & 0.42 & 0.93 & 0.38 \\
 NGC 6522  & 0.51 & 1.61 & 0.52 \\
 NGC 6540  & 4.59 & 6.72 & 0.19 \\
 NGC 6558  & 0.71 & 3.89 & 0.69 \\
 NGC 6652  & 1.64 & 5.89 & 0.56 \\
 NGC 6681  & 0.65 & 4.74 & 0.76 \\
 Palomar 6 & 1.01 & 5.35 & 0.68 \\ \hline
 \end{tabular}\end{center}\end{table}}
%%%%%%%%%%%%%%%%%%%%%%%%%%%%%%%%%%%%%%%%%%%%%%

%%%%%%%%%%%%%%%%%%%%%%%%%%%%%%%%%%%%%%%%%%%%%%%%%%%%%%%
 {\begin{table}[t]                                    %% T~5.
 \caption[]
 {\small\baselineskip=1.0ex
Characteristics of the globular-cluster orbits computed for the
axially symmetric potential (upper rows) and with the addition of
the bar potential (lower rows), taking into account the
measurement uncertainties
 }
 \label{t:err}
 \small\baselineskip=1.0ex
\begin{center}\begin{tabular}{|l|c|c|c|c|c|c|c|}\hline
 Cluster & $ a_{min}$ & $ a_{max}$ & $ e$ \\\hline
           & \multicolumn{2}{c|}{kpc} &     \\\hline
 Terzan 1  & 1.39$\pm$0.10 & 3.80$\pm$0.32 & 0.46$\pm$0.04 \\
 Terzan 2  & 0.51$\pm$0.07 & 1.21$\pm$0.25 & 0.40$\pm$0.09 \\
 Terzan 4  & 0.60$\pm$0.11 & 4.44$\pm$1.27 & 0.75$\pm$0.05 \\
 Terzan 9  & 0.44$\pm$0.10 & 0.91$\pm$0.07 & 0.36$\pm$0.08 \\
 NGC 6522  & 0.51$\pm$0.09 & 1.64$\pm$0.35 & 0.52$\pm$0.07 \\
 NGC 6540  & 4.59$\pm$0.07 & 6.71$\pm$0.56 & 0.19$\pm$0.04 \\
 NGC 6558  & 0.70$\pm$0.15 & 3.92$\pm$0.53 & 0.70$\pm$0.03 \\
 NGC 6652  & 1.64$\pm$0.09 & 5.73$\pm$0.75 & 0.55$\pm$0.05 \\
 NGC 6681  & 0.67$\pm$0.13 & 4.92$\pm$0.82 & 0.75$\pm$0.07 \\
 Palomar 6 & 1.00$\pm$0.08 & 5.29$\pm$0.63 & 0.68$\pm$0.03 \\ \hline

 Terzan 1  & 1.42$\pm$0.12 & 3.92$\pm$0.31 & 0.47$\pm$0.03 \\
 Terzan 2  & 0.51$\pm$0.09 & 1.32$\pm$0.24 & 0.44$\pm$0.11 \\
 Terzan 4  & 0.58$\pm$0.11 & 4.65$\pm$1.29 & 0.77$\pm$0.05 \\
 Terzan 9  & 0.46$\pm$0.10 & 0.92$\pm$0.08 & 0.34$\pm$0.07 \\
 NGC 6522  & 0.48$\pm$0.08 & 1.74$\pm$0.41 & 0.56$\pm$0.08 \\
 NGC 6540  & 4.56$\pm$0.09 & 6.87$\pm$0.56 & 0.20$\pm$0.04 \\
 NGC 6558  & 0.72$\pm$0.16 & 4.03$\pm$0.52 & 0.70$\pm$0.04 \\
 NGC 6652  & 1.61$\pm$0.07 & 5.97$\pm$0.75 & 0.57$\pm$0.05 \\
 NGC 6681  & 0.64$\pm$0.14 & 4.89$\pm$0.75 & 0.76$\pm$0.07 \\
 Palomar 6 & 0.98$\pm$0.08 & 5.66$\pm$0.77 & 0.70$\pm$0.03 \\ \hline
 \end{tabular}\end{center}\end{table}}
%%%%%%%%%%%%%%%%%%%%%%%%%%%%%%%%%%%%%%%%%%%%%%

 \section{DATA}
Our main source of data was [6], where both ground observations on
telescopes of the European Southern Observatory and space
observations with the HST (as a first epoch for NGC 6540) were
used to derive the absolute proper motions of selected globular
clusters of the Galactic bulge. The mean difference between the
epochs was 25 yrs.

The proper motions of two globular clusters located near the
Galactic center ($R<4$ kpc) --- NGC 6652 and NGC 6681---were
determined using HST observations. The proper motion of NGC 6652
was obtained in [29] with a mean difference between epochs of
about seven years, and the proper motion of NGC 6681 in [30] with
a mean difference between epochs of about 5.5 yrs.

The input parameters for the globular clusters are presented in
Table 2, whose columns give (1) the name of the cluster, (2)--(3)
the Galactic coordinates $l$ and $b,$ (4)--(5) the proper motions
$\mu_\alpha\cos\delta$ and $\mu_\delta$ in mas yr$^{-1}$, (6) the
heliocentric radial velocity $V_r,$ and (7) the heliocentric
distance $d.$ The initial values of the space velocities $U,V,W$
and $\Pi,\Theta$ are given in Table 3.

%%%%%%%%%%%%%%%%%%%%%%%% FIG.1:
\begin{figure}[p]
{\begin{center}
 \includegraphics[width=1.0\textwidth]{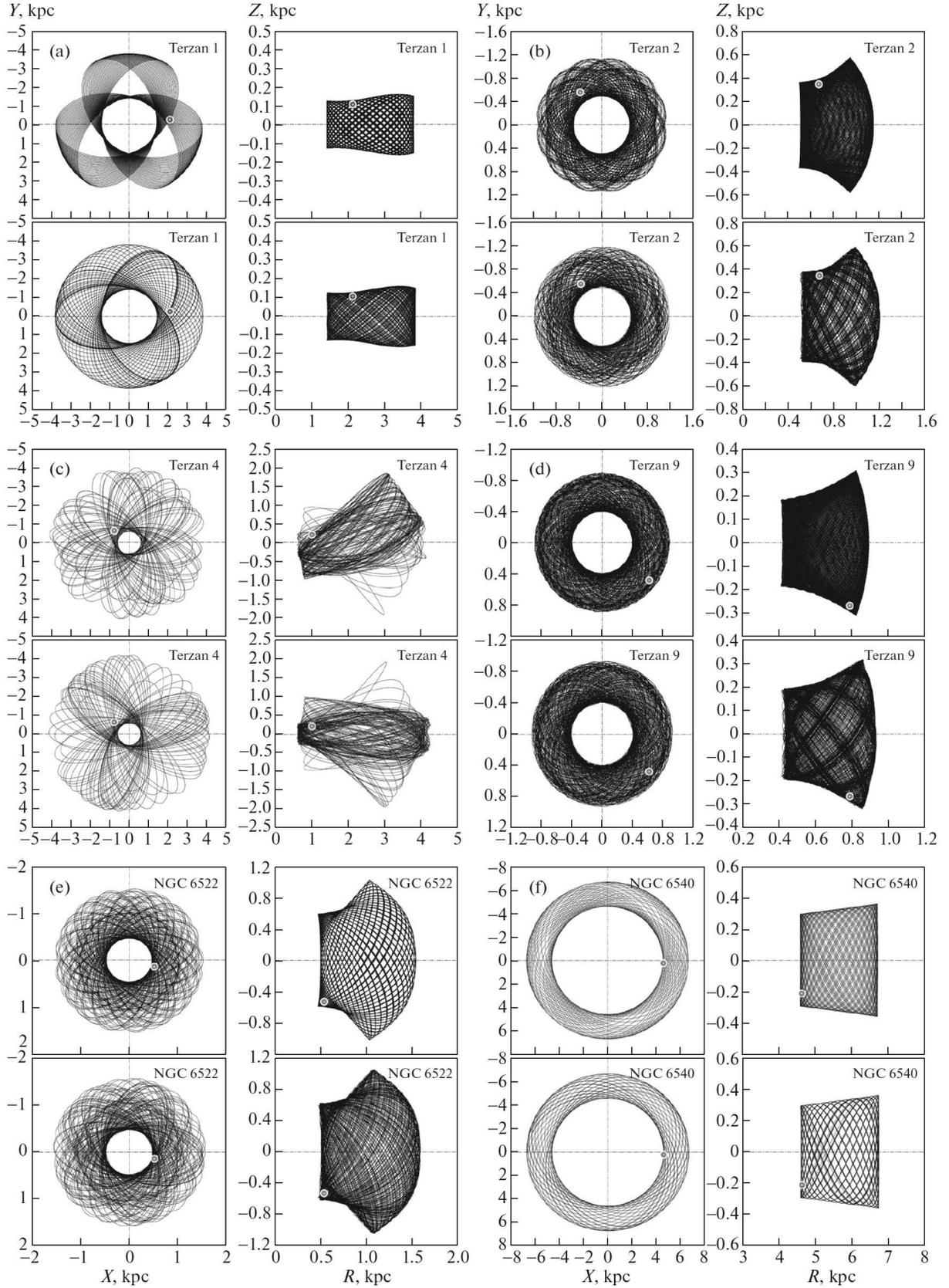}
 \caption{
Galactic orbits of the first six globular clusters from our list
over five billion years in the past. The circle marks the current
position of the cluster.
  } \label{f-1}
\end{center}}
\end{figure}
%%%%%%%%%%%%%%%%%%%%%%%%
%%%%%%%%%%%%%%%%%%%%%%%% FIG.2:
\begin{figure}[p]
{\begin{center}
 \includegraphics[width=1.0\textwidth]{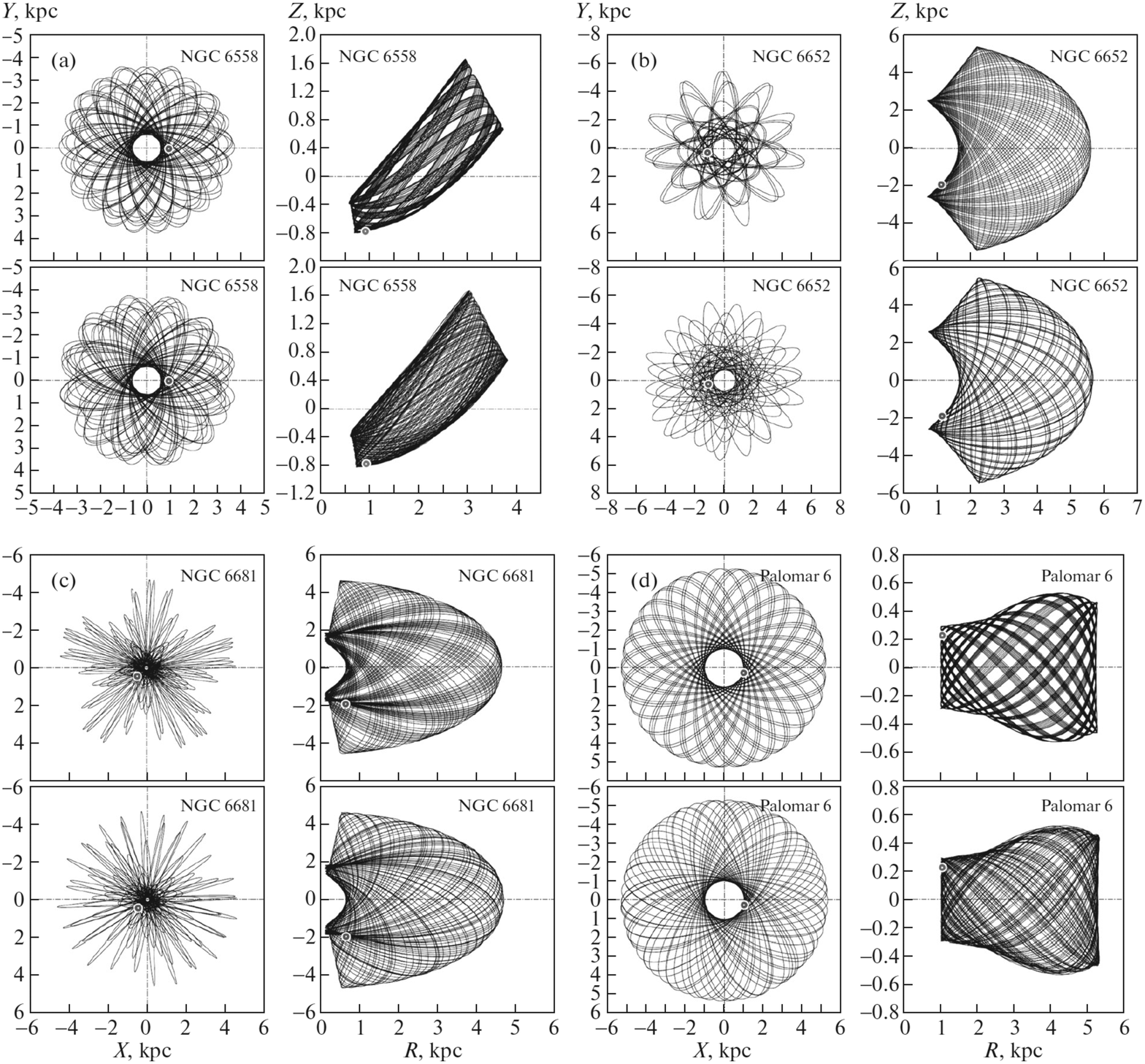}
 \caption{
Same as Fig. 1 for the remaining four globular clusters from our
list.
  } \label{f-2}
\end{center}}
\end{figure}
%%%%%%%%%%%%%%%%%%%%%%%%

 \section{RESULTS AND DISCUSSION}
The Galactic orbits of the globular clusters are presented in
Figs. 1 and 2. Figure 1 presents the orbits of the first six
clusters in our list: (a) Terzan 1, (b) Terzan 2, (c) Terzan 4,
(d) Terzan 9, (e) NGC 6522, and (f) NGC 6540. Figure 2 presents
the orbits for the remaining four clusters: (a) NGC 6558, (b) NGC
6652, (c) NGC 6681, and (d) Palomar 6. Two projections, onto the
$XY$ and $RZ$ planes, are given for each cluster, for the axially
symmetric potential (upper) and with the addition of the bar
potential (lower).

Table 4 presents the perigalactic $a_{min}$ and apogalactic
$a_{max}$ distances and the orbital eccentricities e for both
potential models corresponding to the orbits in Figs. 1 and 2. The
parameters of the globular cluster Terzan 4 varied most strongly
due to the influence of the bar potential: $a_{max}$ increased by
0.12 kpc. For the remaining clusters, the variation in this
parameter was 0.05 kpc or less.

Figures 1 and 2 show that the trajectories of virtually all of the
clusters vary appreciably in the $XY$ plane due to the influence
of the bar. These variations are smaller in the $RZ$ plane, apart
from the cluster Terzan 4, where they are clearly visible in Fig.
1c. The influence of the bar becomes less important with
increasing distance of the cluster from the Galactic center. For
example, the bar exerts virtually no influence for NGC 6540, which
is the most distant cluster from the center ($R=4.6$~kpc, Fig.
1f). A similar picture can be seen in Fig. 1 of [7], where a
different axially symmetric Galactic potential and different bar
potential were used: the bar in [7] was longer ($a_b=3.13$~kpc)
and rotated more rapidly ($\Omega_{bar}=60$ km s$^{-1}$
kpc$^{-1}$) than ours. Figures 7 and 8 of [13] also show that the
influence of the bar is maximum for the orbits of clusters located
closer to the Galactic center, but is present out to
$R\approx6$~kpc.

Uncertainties in the input data for the globular clusters could
have a larger influence than the effect of the bar. For example,
the characteristics $a_{min}=0.96$ kpc, $a_{max}=2.78$ kpc,
$z_{max}=1.43$ kpc, and $e=0.49$ were found for NGC 6522 for the
axially symmetric potential in [7]. A comparison with the
corresponding data for NGC 6522 in Table 4 shows that our values
for $a_{min}$ and $a_{max}$ are nearly half these values, although
the two eccentricities are similar. The orbit parameters for all
the globular clusters we have considered, obtained for both the
axially symmetric potential and with the addition of the bar
potential, taking into account the observational uncertainties,
are presented in Table 5. We determined the mean parameters and
their rms deviations using Monte Carlo simulations, generating 100
independent realizations of random errors in the data for each
object, with these errors having a Gaussian distribution with a
specified rms deviation and zero mean. A comparison of Tables 4
and 5 confirms our conclusion that uncertainties in the data exert
a larger influence than does the bar.

The Galactic orbits of 63 Galactic globular clusters were computed
in [31], using both axially symmetric and non-axially symmetric
Galactic potentials. The non-axially symmetric part of the
potential included contributions from the bar and spiral density
wave. NGC 6522 was included in both [31] and our study. Figure~2
of [31] shows that the orbits (in the $ZR$ plane) constructed for
the axially symmetric potential are very close to those for NGC
6522 shown in Fig.~1. It is interesting, however, that the
appearance of this plane constructed taking into account the joint
influence of the bar and spiral wave differs appreciably from our
results. The orbital characteristics found for NGC 6522 in [31]
using the axially symmetric potential were $a_{min}=0.81$ kpc,
$a_{max}=2.23$~kpc, and $e=0.47$ (these values can also be
compared with the estimates of [7] mentioned above), and those
found using the non-axially symmetric potential were
$a_{min}=0.37$ kpc, $a_{max}=3.87$~kpc, and $e=0.83.$

As can be seen in the upper part of Table 4, the six clusters
Terzan 1, Terzan 2, Terzan 4, Terzan 9, NGC 6522, and NGC 6558 all
have $a_{max}<4.2$~kpc, so that they are always located in the
bulge. It was recently established that the Galactic bulge has an
X-like shape [32,33]. Two hypotheses have been proposed to explain
this. According to the first, the X-like shape is not a physical
property of the bulge, and is a consequence of bimodality in the
distribution in the Hertzsprung–Russell diagram of the giant stars
used to study the bulge [34--36]. Note, however, that pronounced
X-like bulges are observed in other galaxies without any
particular difficulties [37,38]. The second hypothesis is that the
X-like shape has a dynamical nature, and is due to the
characteristic form of stellar orbits in the central region of the
Galaxy [39,40]. Another dynamical approach based on modelling disk
instability whose development leads to the formation of an
X-shaped bulge has also been considered [41]. A final choice of
one or the other hypothesis is not currently possible. In this
connection, the orbits of globular clusters in the bulge
constructed over long time intervals using high-accuracy
observational data are of considerable interest.

In our opinion, the shape of the orbits of the six clusters noted
above provides support for the second dynamical hypothesis for the
origin of the X-shaped form of the bulge. Figure 1 clearly shows
that the orbits of the globular clusters Terzan 2, Terzan 4,
Terzan 9, and NGC 6522 in the $ZR$ plane resemble each other.
Moreover, they resemble a trapezoid with sharply protruding edges
set on its side, so that they trace out an X-like shape upon
mirror reflection relative to the vertical axis.

In spite of the unusual form of the orbit of NGC 6558 ($ZR$ plane,
Fig. 2), an X-like shape is formed upon mirror reflection relative
to the vertical and horizontal axes. The orbit of this cluster can
be considered to be banana-like. Such orbits make an appreciable
contribution to the formation of the X-like shape of the bulge, as
was shown, for example, in the numerical simulations of [39], as
is clearly visible in their Figs.~2 and 3.

The spatial morphology of an X-shaped bar was studied in [41]
using three models with different input parameters. Numerical
solutions were used to trace the evolution of a cloud of particles
under the action of a developing disk instability. The
distribution of the density was constructed for all three models.
It is interesting that the ``boxes'' in the $ZR$ meridianal planes
for all the clusters in Figs.~1 and 2 apart from the distant NGC
6540 are in fairly good agreement with the simulation results of
Li and Shen [41] (see Fig.~5 in [41]).

 \section{CONCLUSION}
We selected high-accuracy measurements of the proper motions,
radial velocities, and distances of ten globular clusters in the
Milky Way for the study considered here. These clusters are
located in the inner part of the Galaxy, at distances of no more
than 5 kpc from the center.

We have constructed their Galactic orbits over a long time
interval, first applying a model with an axially symmetric
gravitational potential for the Galaxy, then with the addition of
the potential of the central bar. Virtually all of the ten
globular clusters display appreciable variations in their
trajectories in the $XY$ plane due to the influence of the bar.
Here, the influence of the bar is mainly manifest through
chaotization of the filling of the annular region bounded by the
values $a_{min}$ and $a_{max}.$ The influence of the bar in the
$RZ$ plane is most clearly visible in the motion of the cluster
Terzan~4. Here, we can see a substantial variation in the
appearance of the corresponding ``box'' (the ``boxes'' coincide
after a longer integration interval).

We have identified a number of clusters that have always been
located within the central bulge over the past five billion years
(no farther than 4 kpc from the Galactic center): Terzan~1,
Terzan~2, Terzan~4, Terzan~9, NGC 6522, and NGC 6558. The shapes
of their orbits are consistent with the hypothesis that the X-like
shape of the bulge has a dynamical nature. The banana-like orbit
of NGC 6558 is of the most interest from this point of view.

 \subsubsection*{ACKNOWLEDGEMENTS}
We thank the referee for useful comments that have helped to
improve this paper. This work was supported by the Basic Research
Program P-7 of the Presidium of the Russian Academy of sciences,
under the subprogram ``Transitional and Explosive Processes in
Astrophysics''.

 \bigskip\subsubsection*{REFERENCES}

 {\small
\quad~ 1. E. Bica, C. Bonatto, B. Barbuy, and S. Ortolani, Astron.
Astrophys. 450, 105 (2006).

 2. R. Leaman, D.A. van den Berg, and J.T. Mendel, Mon. Not. R. Astron. Soc. 436, 122 (2013).

3. G. Battaglia, A. Helmi, H. Morrison, P. Harding, E.W.
Olszewski, M. Mateo, K.C. Freeman, J. Norris, and S. A. Shectman,
Mon. Not. R. Astron. Soc. 364, 433 (2005).

4. Y. Sofue, Publ. Astron. Soc. Jpn. 61, 153 (2009).

5. P. Bhattacharjee, S. Chaudhury, and S. Kundu, Astrophys. J.
785, 63 (2014).

6. L.J. Rossi, S. Ortolani, B. Barbuy, E. Bica, and A. Bonfanti,
Mon. Not. R. Astron. Soc. 450, 3270 (2015).

7. C. Allen, E. Moreno, and B. Pichardo, Astrophys. J. 652, 1150
(2006).

8. C. Allen, E. Moreno, and B. Pichardo, Astrophys. J. 674, 237
(2008).

9. D.V. Putte and M. Cropper, Mon. Not. R. Astron. Soc. 392, 113
(2009).

10. D.I. Dinescu, T.M. Girard, and W.F. van Altena, Astron. J.
117, 1792 (1999).

11. D.I. Casetti-Dinescu, T.M. Girard, D. Herrera, W.F. van
Altena, C.E. L\'opez, and D.J. Castillo, Astron. J. 134, 195
(2007).

12. D.I. Casetti-Dinescu, T.M. Girard, V.I. Korchagin, W.F. van
Altena, and C.E. L\'opez, Astron. J. 140, 1282 (2010).

13. D.I. Casetti-Dinescu, T.M. Girard, L. Jilkov\'a, W.F. van
Altena, F. Podest\'a, and C.E. L\'opez, Astron. J. 146, 33 (2013).

14. T.M. Girard, D.I. Dinescu, W.F. van Altena, I. Platais, D.C.
Monet, and C.E. L\'opez, Astron. J. 127, 3060 (2004).

15. The HIPPARCOS and Tycho Catalogues, ESA SP--1200 (1997).

16. A.K. Dambis, Astron. Astrophys. Trans. 25, 185 (2006).

17. N. Zacharias, S.E. Urban, M.I. Zacharias, G. L. Wycoff,D.M.
Hall, M.E. Germain, E.R. Holdenried, and L. Winter, Astron. J.
127, 3043 (2004).

18. N.V. Kharchenko, A.E. Piskunov, E. Schilbach, S. R\"oser, and
R.-D. Scholz, Astron. Astrophys. 558, A53 (2013).

19. S. R\"oser,M. Demleitner, and E. Schilbach, Astron. J. 139,
2440 (2010).

20. P.N. Fedorov, V.S. Akhmetov, and V.V. Bobylev, Mon. Not. R.
Astron. Soc. 416, 403 (2011).

21. J.J. Vickers, S. R\"oser, and E.K. Grebel, Astron. J. 151, 99
(2016).

22. A.T. Bajkova and V.V. Bobylev, Astron. Lett. 42, 567 (2016).

23. M. Miyamoto and R. Nagai, Publ. Astron. Soc. Jpn. 27, 533
(1975).

24. J.F. Navarro, C.S. Frenk, and S.D.M. White, Astrophys. J. 490,
493 (1997).

25. J. Palou$\breve{s}$, B. Jungwiert, J.~Kopeck\'y, Astron.
Astrophys. 274, 189 (1993).

26. V.V. Bobylev and A.T. Bajkova, Astron. Lett. 42, 228 (2016).

27. A. Irrgang, B. Wilcox, E. Tucker, and L. Schiefelbein, Astron.
Astrophys. 549, 137 (2013).

28. R. Sch\"onrich, J. Binney, and W. Dehnen, Mon. Not. R. Astron.
Soc. 403, 1829 (2010).

29. S.T. Sohn, R.P. van der Marel, J.L. Carlin, S.R. Majewski, N.
Kallivayalil, D.R. Law, J. Anderson, and M.H. Siegel, Astrophys.
J. 803, 56 (2015).

30. D. Massari, A. Bellini, F.R. Ferraro, R.P. van der Marel, J.
Anderson, E. Dalessandro, and B. Lanzoni, Astrophys. J. 779, 81
(2013).

31. E. Moreno, B. Pichardo, and H. Vel\'azquez, Astrophys. J. 793,
110 (2014).

 32. O.A. Gonzalez and D. Gadotti, Astrophys. Space Sci. Lib. 418, 199 (2016).

 33. M. Ness and D. Lang, Astron. J. 152, 14 (2016).

34. A. McWilliam and M. Zoccali, Astrophys. J. 724, 1491 (2010).

35. D.M. Nataf, A. Udalski, J. Skowron, M.K. Szyma\'nski, M.
Kubiak, G. Pietrzy\'nski, I. Soszy\'nski, K. Ulaczyk, et al., Mon.
Not. R. Astron. Soc. 447, 1535 (2015).

36. Y.-W. Lee, S.-J. Joo, and C. Chung, Mon. Not. R. Astron. Soc.
453, 3906 (2015).

37. E. Laurikainen, H. Salo, E. Athanassoula, A. Bosma, and M.
Herrera-Endoqui, Mon. Not. R. Astron. Soc. 444, 80 (2014).

38. E. Laurikainen and H. Salo, Astrophys. Space Sci. Lib. 418, 77
(2016).

39. M. Portail, C. Wegg, and O. Gerhard, Mon. Not. R. Astron. Soc.
450, 66 (2015).

40. E. Athanassoula, Astrophys. Space Sci. Lib. 418, 391 (2016).

41. Z.-Y. Li and J. Shen, Astrophys. J. 815, 20 (2015). }

 \end{document}